\documentclass[aps,prd,showpacs,nofootinbib]{revtex4}
\usepackage{bbm}
\usepackage{mathrsfs}
\usepackage{epsfig,psfrag}
\usepackage{amsmath,amsfonts,amssymb}
\usepackage[usenames]{color}
\usepackage{bm}
\usepackage{ulem}
\newcommand{\tr}{\text{tr }}

\begin{document}

\title{Variable Phase $S$-Matrix Calculations for
Asymmetric Potentials and Dielectrics}
 
\author{Aden Forrow}
\email{aforrow@middlebury.edu}
\affiliation{Department of Physics,
Middlebury College,
Middlebury, VT 05753  USA}

\author{Noah Graham}
\email{ngraham@middlebury.edu}
\affiliation{Department of Physics,
Middlebury College,
Middlebury, VT 05753  USA}

\begin{abstract}
Motivated by recently developed techniques making it possible
to compute Casimir energies for any object whose scattering $S$-matrix
(or, equivalently, $T$-matrix)
is available, we develop a variable phase method to compute the
$S$-matrix for localized but asymmetric sources.  Starting from the
case of scalar potential scattering, we develop a combined
inward/outward integration algorithm that is numerically efficient and
extends robustly to imaginary wave number.  We then extend these
results to electromagnetic scattering from a position-dependent
dielectric.  This case requires additional modifications to
disentangle the transverse and longitudinal modes.
\end{abstract}

\maketitle

\section{Introduction}

Scattering theory \cite{Newton02,Chadan} is an invaluable tool for
investigating a wide range of physical systems.  Far away from a
system that is localized in space, one can express solutions to the
wave equation as free incoming and outgoing partial waves.  In this
partial wave basis, the scattering $S$-matrix then gives the amplitude
and phase of outgoing waves reflected from the system in terms of a
given amplitude and phase of incoming waves.

One of the many applications of scattering theory arises in
calculating Casimir forces.  While the connection between Casimir
forces and scattering amplitudes has long been understood in planar
systems \cite{Kats77,Jaekel91}, only recently have techniques been
developed in which the Casimir force is expressed in terms of the
$S$-matrix  (or, equivalently, $T$-matrix) for general geometries
\cite{spheres,scalar,universal}.  In this approach, the $S$-matrix
encodes the effects of quantum fluctuations on a single object, while
universal translation matrices, obtained from the free Green's
function, encode the objects' relative positions and orientations.
This decomposition provides a concrete implementation of the ``TGTG''
representation of the Casimir energy in terms of scattering transition
operators and free Green's functions \cite{Kenneth06}.  The $S$-matrix
is also a key ingredient in Casimir calculations of quantum
corrections to soliton energies and charges \cite{Graham09}.  These
calculations take advantage of the relationship between the $S$-matrix
and the change in the continuum density of states,
\begin{equation}
\Delta \rho(k) = \tr \frac{1}{\pi}\frac{d}{dk} 
\left(\frac{1}{2 i} \log \hat S_k \right) \,,
\end{equation}
where the eigenvalues of the matrix in parentheses are the scattering
phase shifts.

A standard approach to finding the exact electromagnetic $S$-matrix for
dielectric objects involves integrating the vector solutions
of the Helmholtz equation in dielectric media over the object's surface
\cite{Waterman65,Waterman71}.  A variety of subsequent techniques have
obtained a wide range of analytic and numerical results
\cite{T-matrix-refs}.  In many cases of practical interest, one can
obtain approximate results valid in appropriate limits, such as large
or small values of the wavelength or partial wave number.  Because the
Casimir calculation involves summing over all fluctuating modes,
however, suitable approximations are often not available.

In Casimir problems and other applications of scattering theory, one
frequently considers objects with sufficient symmetry that the problem
separates and the $S$-matrix is diagonal.  For cases where the
resulting ordinary differential equation cannot be solved
analytically, the variable phase method \cite{variable,Graham09}
provides an efficient numerical algorithm.  In particular, it allows
one to solve for the $S$-matrix as an initial value ODE, rather
than a boundary value problem.  Here we extend the variable
phase method to compute the $S$-matrix in situations without any
symmetry assumptions.  We begin with the case of a scalar potential,
as arises in quantum mechanical potential scattering, which we then
generalize to the case of electromagnetism with a position-dependent
dielectric.  Our approach can provide a middle ground between analytic
results and fully general numerical calculations \cite{Johnson}.  For
dielectrics, our work extends the results found in
Ref.~\cite{Johnson99}, which treats the case of a spherically
symmetric but $r$-dependent dielectric.  As we will see, the
asymmetric case introduces additional complications, because one can
no longer rely on the channel decomposition to separate transverse and
longitudinal modes.  We also introduce a combined inward/outward
integration algorithm, which makes use of the Wronskian of the regular
and outgoing solutions, to ensure the stability of the numerical
calculation for imaginary wave number $k=i\kappa$.

\section{Helmholtz Scattering}

We begin by considering scattering of waves obeying the scalar Helmholtz
equation, as would arise in a typical quantum mechanics problem.  This
calculation generalizes straightforwardly to the vector Helmholtz
equation, as we show in this section.  Additional formalism is needed
for the case of Maxwell scattering, however, so we postpone that case
to the next section.

\subsection{Variable Phase Approach:  Outgoing Wave}

We start from the Helmholtz equation in three dimensions
\begin{equation}
-\nabla^2 \psi_k(\bm{r}) + V(\bm{r}) \psi_k(\bm{r}) = k^2 \psi_k(\bm{r}) \,,
\end{equation}
where the potential $V(\bm{r})$ is localized in a region around the
origin.  This equation describes, for example, ordinary
quantum-mechanical scattering of the scalar wavefunction
$\psi_k(\bm{r})$ from a localized potential.  Since each $k$ value is
treated separately, $V(\bm{r})$ can also be $k$-dependent, though we
do not indicate this possibility explicitly.  We expand both the
solution $\psi_k(\bm{r})$ and the potential $V(\bm{r})$ using a Fourier
series in the angular variables,
\begin{equation}
\psi_k(\bm{r}) = \sum_{\ell=0}^\infty  \sum_{m=-\ell}^\ell
\frac{1}{r} \psi_{\ell m,k}(r) Y_{\ell}^{m}(\theta, \phi)
\quad \hbox{and}\quad
V(\bm{r}) =  \sum_{\ell'=0}^\infty  \sum_{m'=-\ell'}^{\ell'}
V_{\ell' m'}(r) Y_{\ell'}^{m'}(\theta, \phi) \,,
\end{equation}
to obtain
\begin{equation}
\sum_{\ell m}  Y_{\ell}^{m}(\theta, \phi)
\left(-\frac{\partial^2}{\partial r^2} + \frac{\ell (\ell+1)}{r^2}
- k^2\right) \psi_{\ell m,k}(r)
+ \sum_{\ell m} \psi_{\ell m,k}(r) Y_{\ell}^{m}(\theta, \phi)
\sum_{\ell' m'} V_{\ell' m'}(r) Y_{\ell'}^{m'}(\theta, \phi) = 0\,.
\end{equation}
Next, we multiply both sides by $Y_{\ell''}^{m''}(\theta, \phi)^\ast =
(-1)^{m''} Y_{\ell''}^{-m''}(\theta, \phi)$
and integrate over solid angle.  The last term becomes a convolution,
which mixes angular momentum channels.  We obtain
\begin{equation}
\left(-\frac{\partial^2}{\partial r^2} + \frac{\ell'' (\ell''+1)}{r^2}
- k^2\right) \psi_{\ell'' m'',k}(r)
+ \sum_{\ell m} \left(\sum_{\ell' m'} V_{\ell' m'}(r)
Z_{\ell \ell' \ell''}^{mm'm''}\right) \psi_{\ell m,k}(r) = 0 \,,
\label{eqn:schro}
\end{equation}
where the integral identity
\begin{eqnarray}
\int_0^\pi \sin \theta d\theta \int_0^{2\pi} d\phi \,
Y_{l}^{m}(\theta,\phi)Y_{l'}^{m'}(\theta,\phi)Y_{l''}^{m''}(\theta,\phi) = 
\sqrt{\frac{(2\ell+1)(2\ell'+1)(2\ell''+1)}{4\pi}}
\begin{pmatrix}
\ell & \ell' & \ell'' \cr 0 & 0 & 0 \\
\end{pmatrix}
\begin{pmatrix}
\ell & \ell' & \ell'' \cr m & m' & m''
\end{pmatrix}
\label{eqn:scalarwigner}
\end{eqnarray}
allows us to express $Z_{\ell \ell' \ell''}^{mm'm''}$ in terms of
$3j$-symbols as
\begin{equation}
Z_{\ell \ell' \ell''}^{mm'm''} = (-1)^{m''}
\sqrt{\frac{(2 \ell + 1)(2 \ell' + 1)(2 \ell'' + 1)}{4 \pi}}
\begin{pmatrix}
\ell & \ell' & \ell'' \cr 0 & 0 & 0 \\
\end{pmatrix}
\begin{pmatrix}
\ell & \ell' & \ell'' \cr m & m' & -m''
\end{pmatrix} \,.
\label{eqn:scalarcoupling}
\end{equation}
In the absence of a potential, the regular and outgoing
solutions for $\psi_{\ell m,k}(r)$ are given in terms of
spherical Bessel and spherical Hankel functions by $kr j_\ell(kr)$ and
$kr h^{(1)}_\ell(kr)$, respectively.

Since the scattering channels will mix for a nonspherical potential,
we will want to consider all incoming waves together.  To do so, we
rewrite Eq.~(\ref{eqn:schro}) as a matrix differential equation,
\begin{equation}
\left(-\frac{\partial^2}{\partial r^2} + \frac{\hat L^2}{r^2}
- k^2\right) \hat \psi_k(r)
+ \hat V(r) \hat \psi_k(r) = 0 \,,
\end{equation}
where hat indicates a matrix indexed by the angular momentum
indices $\ell$ and $m$ (so that both $\ell$ and $m$ are combined into
a single matrix index), $\hat L^2$ is a
diagonal matrix with $\ell(\ell+1)$ on the diagonal, and 
$\hat V(r)$ is the matrix in parentheses in the second term of
Eq.~(\ref{eqn:schro}).  

We begin by considering the solution to this equation with outgoing
wave boundary conditions, which we parameterize as
\begin{equation}
\hat F_k(r) = \hat G_k(r) \hat W(kr) \,,
\label{eqn:outgoing}
\end{equation}
where $\hat W(x)$ is a diagonal matrix with the free outgoing wave
solutions $x h_\ell^{(1)}(x)$ on the diagonal.  The Helmholtz equation 
for $\hat F_k(r)$ then translates into an ordinary differential equation
for the matrix $\hat G_k(r)$,
\begin{equation}
-\hat G_k''(r) - 2 \hat G_k'(r) 
\left(\frac{\partial}{\partial r} \log \hat W(kr)\right)
+ \frac{1}{r^2} [\hat L^2, \hat G_k(r)] + \hat V(r) \hat G_k(r) = 0 \,,
\label{eqn:g}
\end{equation}
where prime denotes derivative with respect to $r$ and we have used
the fact that the free solution obeys
\begin{equation}
-\hat W''(kr) + \frac{\hat L^2}{r^2} \hat W(kr) = k^2 \hat W(kr) \,,
\label{eqn:free}
\end{equation}
and then multiplied from the right by $\hat W^{-1}(kr)$.  By the outgoing
wave boundary condition, we have $\hat G_k(\infty) = \hat 1$ and $\hat
G_k'(\infty) = \hat 0$, where $\hat 1$ and $\hat 0$ are the identity
and zero matrices respectively.  These results provide the necessary
initial values for integrating Eq.~(\ref{eqn:g}) inward from infinity
to the origin.

To define the $S$-matrix, we combine the solutions with $k$ and
$-k$ (or, equivalently, the outgoing wave solution and its
conjugate, the incoming wave solution) to form the physical wave
function,
\begin{equation}
\hat \psi_k(r) = -\hat G_{-k}(r) \hat W(-kr) \hat M + 
\hat G_k(r) \hat W(kr) \hat S_k(k) \,,
\end{equation}
where $\hat M$ is a diagonal matrix with $(-1)^\ell$ on the diagonal.
We then find the $S$-matrix by the regularity condition at the origin,
which yields
\begin{equation}
\hat S_k=\lim_{r\to 0} \hat W^{-1}(kr) \hat G_k^{-1}(r) \hat G_{-k}(r)
\hat W(-kr) \hat M \,.
\label{eqn:Smatrix}
\end{equation}
In many applications it is convenient to work with the $T$-matrix,
which is given by $\hat T_k=\frac{1}{2}(\hat S_k-\hat 1)$.

We can thus find the $S$-matrix numerically, by integrating $\hat
G_k(r)$ in from $r=\infty$ to $r=0$, and similarly for $\hat
G_{-k}(r)$. The combination $\displaystyle \hat W'(x) \hat W^{-1}(x) =
\frac{\partial}{\partial x} \log \hat W(x)$ is easy to calculate
numerically, since it is just a diagonal matrix with rational
functions of $x$ on the diagonal, which can be obtained from a finite
continued fraction expansion \cite[p.\ 241]{numrec}.
The inputs to the calculation are then the ``multipole
moments'' of the potential at each $r$, $V_{\ell m}(r)$.  We
could imagine some simple non-spherical potentials for which these
moments might be particularly easy to find; or, we could specify the
potential explicitly through its representation in this spherical
harmonic basis.

Note that in the ordinary variable phase method,
where the channels separate (so here all matrices would be diagonal),
it is common to write $G_k(r)=e^{i\beta_k(r)}$, which further simplifies
the calculation.  This approach is problematic in the general case,
however, because then $\hat \beta_k(r)$ doesn't commute with its derivatives.

\subsection{Variable Phase Approach: Regular Wave}
\label{sec:reg}

In principle, one could carry out the calculation of the previous
subsection for $k=i\kappa$ to obtain the $S$-matrix on the
imaginary $k$-axis, as  is typically required in Casimir calculations.
In practice, however, this is not possible, because in place of the
oscillating spherical Bessel function $h^{(1)}_\ell(kr)$, we now have
the exponentially decaying modified function $k_\ell(\kappa r)$, which
then grows exponentially as we integrate in from infinity.  As a result,
a direct application of the previous results is hopelessly unstable
numerically, and we will need to introduce some additional formalism
to obtain a useful calculation.

To address this problem, we use an approach developed in
Ref.~\cite{density}, in which we parameterize the regular
solution in a complementary way to what we did for the outgoing
solution in Eq.~(\ref{eqn:outgoing}).  Here it will be convenient to
parameterize the transpose of the regular solution, as
\begin{equation}
\hat \Phi_k(r)^t = \hat W(kr)^{-1} \hat H_k(r)
\end{equation}
(note the reversed order in this decomposition).  We then have
\begin{equation}
-\hat H_k''(r) + 2 \frac{\partial}{\partial r} \left[
\left(\frac{\partial}{\partial r} \log \hat W(kr)\right)
\hat H_k(r)\right]
- \frac{1}{r^2} [\hat L^2, \hat H_k(r)] +  \hat H_k(r) \hat V(r)
 = 0 \,.
\label{eqn:h}
\end{equation}
By the regularity of $\hat \Phi_k(r)^t$ at the origin, we have the
boundary condition $\hat H_k(0) = \hat 0$ and $\hat H_k'(0) = \hat 1$,
where again prime denotes a derivative with respect to $r$.  Starting
from this boundary condition, we can then integrate Eq.~(\ref{eqn:h})
outward from the origin.

This integration also contains instabilities for $k$
imaginary, but what will be useful to us is that they show up in a
complementary region:  The integration of $\hat G_k(r)$ blows up for 
$r \to 0$, while the integration of $\hat H_k(r)$ blows up for 
$r\to \infty$.  We can make use of this complementarity by considering
the Wronskian of our two solutions \cite[p.\ 465]{Newton02},
\begin{eqnarray}
\left.{\cal W}_k\right\vert_r = {\cal W}[\hat \Phi_k(r)^t, \hat F_k(r)] &=& 
\hat \Phi_k(r)^t \left(\frac{\partial}{\partial r} \hat F_k(r)\right) -
\left(\frac{\partial}{\partial r} \hat \Phi_k(r)^t\right) \hat F_k(r) 
\cr
&=& 
\left(\hat W(kr)^{-1} \hat H_k(r) \right)
\left(\hat G_k(r) \hat W'(kr) + \hat G_k'(r) \hat W(kr)\right) \cr
&& - 
\left(\frac{\partial}{\partial r}\left(\hat W(kr)^{-1}\right) \hat H_k(r) 
+  \hat W(kr)^{-1} \hat H_k'(r) \right)
\left(\hat G_k(r) \hat W(kr)\right) \cr
&=& \hat W(kr)^{-1} \left[\hat H_k(r) 
\left(\hat G_k(r) \left(\frac{\partial}{\partial r} \log \hat W(kr) \right)
+ \hat G_k'(r)\right) -
\right. \cr && \left.
\left(\hat H_k'(r) - \left(\frac{\partial}{\partial r} 
\log \hat W(kr)\right)
\hat H_k(r)\right) \hat G_k(r)\right] \hat W(kr)\,,
\label{eqn:Wronk}
\end{eqnarray}
which is independent of $r$.  By the boundary conditions on  $\hat
G_k(r)$ and $\hat H_k(r)$, we also have
\begin{equation}
\lim_{r\to 0} {\cal W}[\hat \Phi_k(r)^t, \hat F_k(r)] =
\lim_{r\to 0} \left[-
\hat W(kr)^{-1} \hat G_k(r) \hat W(kr) \right] \,.
\end{equation}
Thus, at any $r$,
\begin{equation}
{\cal W}[\hat \Phi_k(r)^t, \hat F_k(r)] = \lim_{r\to 0} \left[-
\hat W(kr)^{-1} \hat G_k(r) \hat W(kr) \right] \,.
\label{eqn:Wronk2}
\end{equation}
But the right-hand side of this equation gives the quantity we need 
to calculate the $S$-matrix from Eq.~(\ref{eqn:Smatrix}).  So our
strategy will be to pick an intermediate radius $r_0$ and integrate
both $\hat G_k(r)$ in from $r=\infty$ to $r=r_0$ and $\hat H_k(r)$ out from
$r=0$ to $r=r_0$.  Then we can evaluate the Wronskian in
Eq.~(\ref{eqn:Wronk}) at $r=r_0$ and use it to obtain the right-hand
side of Eq.~(\ref{eqn:Wronk2}), which is what we need to find the
$S$-matrix.  This procedure will continue to be stable even when $k$
is imaginary (with either sign of its imaginary part --- and we will
need both signs to compute the $S$-matrix).

We thus obtain
\begin{equation}
\hat S_k =
\left({\cal W}[\hat \Phi_k(r)^t, \hat F_k(r)]^{-1}\right\vert_{r=r_0}
\left(\hat W(kr)^{-1} \hat W(-kr) \right\vert_{r\to 0}
\left({\cal W}[\hat \Phi_{-k}(r)^t, \hat F_{-k}(r)]\right\vert_{r=r_0}
\hat M \,.
\label{eqn:SWronk}
\end{equation}
This expression is now suitable for numerical evaluation.

\subsection{Vector Helmholtz Equation}

We next generalize this calculation to the vector Helmholtz equation,
\begin{equation}
-\nabla^2 \bm{\psi}_k(\bm{r}) + V(\bm{r}) \bm{\psi}_k(\bm{r}) =  k^2
 \bm{\psi}_k(\bm{r}) \,,
\end{equation}
where our wavefunction is now a three-component vector
$\bm{\psi}_k(\bm{r})$.  Our eventual goal is to study electromagnetic
scattering, which will require significant additional modifications of
this approach to disentangle the transverse and longitudinal modes.
In contrast, the generalization to the vector Helmholtz equation is
relatively straightforward, requiring only that we establish
corresponding definitions and identities appropriate to the vector
case, which we take from Ref.\ \cite{Varshalovich:1988ye}.

We begin by defining the three vector spherical harmonics
for each value of $j=0,1,2,3\ldots$ and $m=-j\ldots j$, 
\begin{equation}
\bm{Y}^{\ell}_{jm} =
\sum_{\sigma=-1}^{+1} \sum_{m'=-\ell}^\ell
C_{\ell m' 1 \sigma}^{j m} Y_\ell^{m'}(\theta,\phi) \bm{e}_\sigma \,,
\end{equation}
where $\ell=j,j\pm 1$ for our three vector spherical harmonics,
$C_{\ell m 1 \sigma}^{j m}$ is a Clebsch-Gordan
coefficient, and the spherical basis vectors are
\begin{eqnarray}
\bm{e}_1 &=& -\frac{e^{i\phi}}{\sqrt{2}} 
\left(\sin \theta \, \bm{\hat{r}} + \cos \theta \, \bm{\hat{\theta}}
+ i \, \bm{\hat \phi} \right)
= -\frac{1}{\sqrt{2}}(\bm{\hat{x}} + i \bm{\hat{y}})\cr
\bm{e}_{0} &=& \cos \theta \, \bm{\hat{r}} - \sin \theta
\, \bm{\hat{\theta}} = \bm{\hat z} \cr
\bm{e}_{-1} &=& \frac{e^{-i\phi}}{\sqrt{2}} 
\left(\sin \theta \, \bm{\hat{r}} + \cos \theta \, \bm{\hat{\theta}}
- i \, \bm{\hat \phi} \right)
=\frac{1}{\sqrt{2}} (\bm{\hat{x}} - i \bm{\hat{y}}) \,.
\end{eqnarray}
  For $j=0$, we have only the case $\ell=1$.  This representation
effectively couples the orbital angular momentum $\ell$ to the $s=1$
spin angular momentum associated with the vector index.  We can then
decompose $\bm{\psi}(\bm{r})$ as
\begin{equation}
\bm{\psi}_k(\bm{r}) = 
\sum_{j=0}^\infty \sum_{\ell = |j-1|}^{j+1} \sum_{m=-j}^j
\frac{1}{r} \psi_{j \ell m,k}(r) 
\bm{Y}^{\ell}_{jm}(\theta,\phi) \,.
\end{equation}
The free outgoing wave solutions to the vector Helmholtz equation are
then $kr h^{(1)}_\ell(kr) \bm{Y}^\ell_{j m}(\theta, \phi)$.
The vector spherical harmonics are orthonormal in the usual way,
\begin{equation}
\int_0^\pi \sin \theta d\theta \int_0^{2\pi} d\phi \,
\bm{Y}^{\ell_1}_{j_1 m_1}(\theta, \phi)^\ast \cdot
\bm{Y}^{\ell_2}_{j_2 m_2}(\theta, \phi)
= \delta_{j_1 j_2} \delta_{\ell_1 \ell_2} \delta_{m_1 m_2} 
\end{equation}
and under complex conjugation they transform as $\bm{Y}^{\ell}_{j
m}(\theta, \phi)^\ast = (-1)^{j+\ell+m+1} \bm{Y}^{\ell}_{j -m}(\theta,
\phi)$.

We can now use the basis of free spherical vector waves
to set up the variable phase calculation in the same way as in the
scalar case.  In place of Eq.~(\ref{eqn:scalarwigner}), we will need
the integral over solid angle of the dot product of two vector
spherical harmonics multiplied by a third ordinary spherical harmonic
(since the potential is still expanded in terms of ordinary spherical
harmonics), which is given in terms of the $6j$-symbol and
Clebsch-Gordan coefficients as
\begin{eqnarray}
\int_0^\pi \sin \theta d\theta \int_0^{2\pi} d\phi \,
\bm{Y}^{\ell_1}_{j_1 m_1}(\theta, \phi) \cdot
\bm{Y}^{\ell_2}_{j_2 m_2}(\theta, \phi)
Y_\ell^{m}(\theta,\phi) =
\hspace{3in} \cr
(-1)^{j_2+\ell_1 + \ell} (-1)^m
\sqrt{\frac{(2j_1+1)(2j_2+1)(2\ell_1+1)(2\ell_2+1)}{4 \pi (2\ell+1)}}
\begin{Bmatrix}
\ell_1 & \ell_2 & \ell \cr j_2 & j_1 & 1
\end{Bmatrix}
C_{\ell_1 0 \ell_2 0}^{\ell 0} C_{j_1 m_1 j_2 m_2}^{\ell -m} \,.
\label{eqn:6j}
\end{eqnarray}
In place of Eq.~(\ref{eqn:scalarcoupling}), we then have the coupling
between channels
\begin{equation}
Z_{j \ell \ell' j'' \ell''}^{mm'm''} = 
(-1)^{\ell'' + \ell' + \ell + m'' + m' + 1}
\sqrt{\frac{(2j+1)(2j''+1)(2\ell+1)(2\ell''+1)}{4 \pi (2\ell'+1)}}
\begin{Bmatrix}
\ell & \ell'' & \ell' \cr j'' & j & 1
\end{Bmatrix}
C_{\ell 0 \ell'' 0}^{\ell' 0} C_{j m j'' -m''}^{\ell' -m'} \,.
\end{equation}
With this modification, the calculation of the $S$-matrix for the
vector Helmholtz equation proceeds analogously to the scalar case.

\section{Generalization to Maxwell's Equations}

To generalize to the case of electromagnetic scattering, we 
consider a linear, spatially-dependent dielectric with no free charge.
The permittivity $\epsilon(\bm{r})$ goes to one at large distances.
We will treat each frequency $\omega = c\sqrt{k^2}$ separately, so our
formalism can easily incorporate frequency dependence in
$\epsilon(\bm{r})$, though as in the scalar case we do not indicate
this possibility explicitly.  The permittivity can also include an
imaginary part, representing dissipation.  We are interested in
solutions to the Maxwell wave equation
\begin{equation}
\nabla \times \nabla \times \bm{E}_k(\bm{r}) = 
k^2 \epsilon(\bm{r}) \bm{E}_k(\bm{r}) \,,
\label{eqn:Maxwell}
\end{equation}
for $k\neq 0$.  Such solutions automatically obey Gauss's law
$\nabla \cdot \bm{D}_k(\bm{r}) = 0$, where 
$\bm{D}_k(\bm{r}) = \epsilon(\bm{r}) \bm{E}_k(\bm{r})$.
However, the solutions to this equation do not span
the full space of vector functions, because in addition to these transverse
solutions there also exist longitudinal solutions, which can be
written as the gradient of a scalar function and therefore solve
Eq.~(\ref{eqn:Maxwell}) with $k=0$.  This situation is problematic for
the variable phase approach (in which we consider each $k$ separately),
because it implies that the matrix coefficient of the second
derivative operator for fixed nonzero $k$ will not be invertible,
leading to an implicit differential-algebraic equation.  We thus
consider a modified equation that allows us to find the $S$-matrix for
the transverse modes while avoiding this problem.

\subsection{Transverse and Longitudinal Modes}

To motivate our approach, we review a common method for solving
the Maxwell wave equation in free space (or within a dielectric with
constant permittivity), which is to replace the curl-curl operator
$\nabla \times \nabla \times$ by minus the Helmholtz operator
$-\nabla^2$. These operators commute, so they share the same
eigenstates, and when acting on the transverse states, they share the
same eigenvalues.  (Recall that 
$-\nabla^2 \bm{E}_k(\bm{r}) = \nabla \times \nabla \times \bm{E}_k(\bm{r})
- \nabla \left(\nabla \cdot \bm{E}_k(\bm{r}) \right)$, 
where for transverse modes in empty space
$\nabla \cdot \bm{E}_k(\bm{r})=0$ by Gauss's Law.)
However, when acting on the longitudinal modes, the
eigenvalue of $-\nabla^2$ is the usual value of $k^2$ associated with
a mode with wave number $k$, rather than zero.  Once all the solutions
to the Helmholtz equation have been identified, it then is usually
straightforward to discard the longitudinal modes and keep only the
transverse modes.

We now generalize this procedure for the case of a position-dependent
dielectric.  We first rewrite Eq.~(\ref{eqn:Maxwell}) in operator form
as
\begin{equation}
\left(\frac{1}{\epsilon(\bm{r})}
\nabla \times \nabla \times \ldots
\right) \bm{E}_k(\bm{r}) = k^2 \bm{E}_k(\bm{r}) \,,
\label{eqn:Maxwellop}
\end{equation}
where $\ldots$ represents the argument of the operator.  We then
define the generalized Helmholtz operator as
\begin{equation}
\left(\frac{1}{\epsilon(\bm{r})}
\nabla \times \nabla \times\ldots - 
\nabla [\nabla \cdot (\epsilon(\bm{r})\cdot \ldots)] \right)
\bm{E}_k(\bm{r}) = k^2 \bm{E}_k(\bm{r}) \,,
\label{eqn:generalizedHelmholtzop}
\end{equation}
which gives the same situation as in the free case:  The operators in
Eqs.~(\ref{eqn:Maxwellop}) and (\ref{eqn:generalizedHelmholtzop})
commute and share the same eigenstates.  For the transverse modes, they
share the same eigenvalues as well, but for the longitudinal modes,
the eigenvalue of Eq.~(\ref{eqn:Maxwellop}) is $k^2=0$, while the
eigenvalue of Eq.~(\ref{eqn:generalizedHelmholtzop}) is the usual
nonzero value of $k^2$ associated with a mode of wave number $k$.
We note that this approach would continue to work in the presence of
a nontrivial permeability $\mu(\bm{r})$, with the only change being
that $\nabla \times \nabla \times$ is replaced by $\nabla \times
\frac{1}{\mu(\bm{r})} \nabla \times$\,.

We will thus solve for the $S$-matrix associated with the wave
equation
\begin{equation}
\nabla \times \nabla \times \bm{E}_k(\bm{r}) - \epsilon(\bm{r})
\nabla [\nabla \cdot (\epsilon(\bm{r})\cdot \bm{E}_k(\bm{r}))]
 = k^2 \epsilon(\bm{r}) \bm{E}_k(\bm{r}) \,.
\label{eqn:generalizedHelmholtz}
\end{equation}
Again, we decompose both the solution and the source in the appropriate
spherical harmonic basis,
\begin{equation}
\bm{E}_k(\bm{r}) =
\sum_{j=0}^\infty \sum_{\ell = |j-1|}^{j+1} \sum_{m=-j}^j
\frac{1}{r} E_{j \ell m,k}(r) 
\bm{Y}^{\ell}_{jm}(\theta,\phi)
\qquad
\epsilon(\bm{r}) = \sum_{\ell'=0}^\infty  \sum_{m'=-\ell'}^{\ell'} 
\epsilon_{\ell' m'}(r) Y_{\ell'}^{m'}(\theta, \phi) \,,
\end{equation}
where $\epsilon_{\ell m}(r)$ goes to $\sqrt{4\pi} 
\delta_{\ell 0} \delta_{m 0}$ at large $r$.  As above, we denote the
matrix outgoing wave solution, written in the vector spherical
harmonic basis, by $\hat F_k(r)$.   We then substitute this expression
into Eq.~(\ref{eqn:generalizedHelmholtz}) and carry out the vector
spherical harmonic algebra symbolically in Mathematica, using the
identities in Appendix \ref{appa} to implement the differential
operators and Eq.~(\ref{eqn:6j}) to carry out the convolution involved
in multiplying by $\epsilon(\bm{r})$.

The result is an equation of the form
\begin{equation}
- \hat d_2(k,r) \hat F_k''(r) + \hat d_1(k,r) \hat F_k'(r) + 
\hat d_0(k,r) \hat F_k(r) = 0 \,,
\end{equation}
where the matrices $\hat d_0(k,r)$, $\hat d_1(k,r)$, and $\hat
d_2(k,r)$ can depend on the dielectric profile and its derivatives, and
prime denotes a derivative with respect to $r$.  The replacement of
Eq.~(\ref{eqn:Maxwellop}) by Eq.~(\ref{eqn:generalizedHelmholtzop})
ensures that $d_2(k,r)$ is an invertible matrix, so we let
$\hat D_1(k,r) =  \left(\hat d_2(k,r)\right)^{-1} \hat d_1(k,r)$
and $\hat D_0(k,r) = 
\left(\hat d_2(k,r)\right)^{-1} \hat d_0(k,r)$ to obtain
\begin{equation}
-\hat F_k''(r) + \hat D_1(k,r) \hat F_k'(r) + \hat D_0(k,r) \hat
F_k(r) = 0 \,.
\label{eqn:EMout}
\end{equation}
Furthermore, since the generalized Helmholtz operator in
Eq.~(\ref{eqn:generalizedHelmholtzop}) approaches the
ordinary Helmholtz operator as $\epsilon\to 1$, for large $r$ this
equation approaches the ordinary Helmholtz equation, with $\hat
D_1(k,r) = 0$ and $\hat D_0(k,r) = \frac{\hat L^2}{r^2} -k^2$.  Again
using Mathematica to carry out the symbolic algebra, we parameterize the
outgoing solution by $\hat F_k(r) = \hat G_k(r) \hat W(kr)$ and, taking
advantage of the simplifications arising from Eq.~(\ref{eqn:free}),
obtain an ordinary matrix differential equation for $\hat G_k(r)$,
\begin{equation}
-\hat G_k''(r) + \left(\hat D_1(k,r) \hat G_k(r) - 2 \hat G_k'(r)\right) 
\left(\frac{\partial}{\partial r} \log \hat W(kr)\right)
+ \hat D_1(k,r) \hat G_k'(r)
+ \left(\hat D_0(k,r) + k^2\right) \hat G_k(r) - 
\hat G_k(r)\frac{\hat L^2}{r^2} = 0 \,,
\end{equation}
with the boundary conditions $\hat G_k(\infty) = \hat 1$ and
$\hat G_k'(\infty) = \hat 0$.

The solutions to Eq.~(\ref{eqn:generalizedHelmholtz}) include both the
transverse solutions to the Maxwell equation that we are looking for
and the longitudinal modes that we wish to discard.  Because the
$S$-matrix is defined in terms of incoming and outgoing asymptotic
waves, it is straightforward to project out the transverse modes.  In
the free case, the transverse solutions are given by \cite{Biedenharn}
\begin{eqnarray}
\bm{M}_{jm,k}(r,\theta,\phi) &=& z_{j}(kr) 
\bm{Y}^{\ell=j}_{jm}(\theta, \phi) \\
\bm{N}_{jm,k}(r,\theta,\phi) &=& 
-\sqrt{\frac{j+1}{2j+1}} z_{j-1}(kr) 
\bm{Y}^{\ell=j-1}_{jm}(\theta, \phi)
+\sqrt{\frac{j}{2j+1}} z_{j+1}(kr) 
\bm{Y}^{\ell= j+1}_{jm}(\theta, \phi)
\label{eqn:transverse}
\end{eqnarray}
for $j=1,2,3\ldots$, where $z_\ell(kr)$ is the appropriate spherical
Bessel or Hankel function of order $\ell$.  Since we have free
electromagnetic waves far away from the dielectric, by simply
projecting the $S$-matrix onto the subspace spanned by these
transverse solutions at large distances, we obtain the full
electromagnetic $S$-matrix.

\subsection{Inward/Outward Integration in the Maxwell Case}

The presence of first-derivative terms in Eq.~(\ref{eqn:EMout})
necessitates some modifications of the Wronskian analysis that we used
in Sec.~\ref{sec:reg} to obtain the $S$-matrix by combining the
outgoing and regular solutions at an intermediate fitting point.  We
consider the transpose of the regular solution, obeying
\begin{equation}
-\hat \Phi_k''(r)^t -
\left(\hat \Phi_k(r)^t \hat D_1(k,r) \right)'
+ \hat \Phi_k(r)^t \hat D_0(k,r) = 0 \,.
\label{eqn:EMreg}
\end{equation}
which we again parameterize by $\hat \Phi_k(r)^t = \hat
W(kr)^{-1} \hat H_k(r)$.  We obtain the differential equation
\begin{eqnarray}
-\hat H_k''(r) + 
\left(\frac{\partial}{\partial r} \log \hat W(kr)\right)
\left(\hat H_k(r) \hat D_1(k,r) + 2 \hat H_k'(r)\right) 
+ 2 \left(\frac{\partial^2}{\partial r^2} \log \hat W(kr)\right) \hat
H_k(r) \cr
- \hat H_k'(r)\hat D_1(k,r) 
- \hat H_k(r) \hat D_1'(k,r) 
+ \hat H_k(r) \left(\hat D_0(k,r) + k^2\right) - 
\frac{\hat L^2}{r^2} \hat H_k(r) = 0 \,,
\end{eqnarray}
with the boundary conditions $\hat H_k(0) = \hat 0$ and
$\hat H_k'(0) = \hat 1$.  Now the quantity that is independent of $r$
is not the Wronskian but instead
\begin{equation}
\left.\widetilde{\cal W}_k\right\vert_r =
\widetilde{\cal W}[\hat \Phi_k(r)^t, \hat F_k(r)] =
{\cal W}[\hat \Phi_k(r)^t, \hat F_k(r)] - \Phi_k(r)^t 
\hat D_1(k,r) \hat F_k(r) \,.
\label{eqn:Wtilde}
\end{equation}
Because the additional term in Eq.~(\ref{eqn:Wtilde}) vanishes at $r=0$, the
expression for the electromagnetic $S$-matrix in terms of
$\widetilde{\cal W}$ is the same as in Eq.~(\ref{eqn:SWronk}), with
${\cal W}_k\vert_{r=r_0}$ replaced by $\widetilde{\cal W}_k\vert_{r=r_0}$.

\section{Numerical Results}

We have constructed ``proof of concept''
implementations of these calculations using Mathematica, which are 
available from {\tt http://community.middlebury.edu/\~{}ngraham}\,.
This high-level code provides a convenient illustration of our
approach for small- to moderate-scale problems; more extensive
calculations are likely to require lower-level code making use of
parallel linear algebra packages.  In this section we describe sample
calculations that use this code to verify and illustrate our approach.

\subsection{Consistency Checks}
Because some of the calculations we have described are the first of
their kind, not all of our results can be compared with previous work.
Nonetheless, we can verify a variety of complementary aspects of our
calculations against known results or consistency conditions.  In
particular, we can check the following:
\begin{itemize}
\item
For potential scattering with real $V(\bm{r})$ and electromagnetic
scattering with real $\epsilon(\bm{r})$, the $S$-matrix should
be unitary, $\hat S_k^\dagger \hat S_k = \hat 1$, for
real $k$.
\item
For electromagnetic scattering, the $S$-matrix we obtain from solving
Eq.~(\ref{eqn:generalizedHelmholtz}) should commute with projection
onto the asymptotic free transverse modes in
Eq.~(\ref{eqn:transverse}).
\item
For scalar, vector, and electromagnetic scattering, the result of the
inward/outward calculation should be independent of the fitting point
$r_0$.
\item
For a spherical finite square well in the scalar case and a dielectric
sphere in the electromagnetic case, the $S$-matrix is diagonal and can
be found analytically.  For the scalar spherical square well, we have
\cite[p.\ 309]{Newton02}
\begin{equation}
S_{k,\ell} = 
-\frac{q h_\ell^{(2)}(ka) j_\ell'(qa)  - k h_\ell^{(2)\prime}(ka) j_\ell(qa)}
{q h_\ell^{(1)}(ka) j_\ell'(qa)  - k h_\ell^{(1)\prime}(ka) j_\ell(qa)} \,,
\end{equation}
where the potential is
\begin{equation}
V(\bm{r}) = \left\{ \begin{array}{l@{\quad}l}
V_0 & r<a \cr
0   & r>a
\end{array} \right. \,,
\end{equation}
and $q=\sqrt{k^2+V_0}$.  For the dielectric sphere, we have
\cite[p.\ 49]{Newton02}
\begin{equation}
S_{k,\ell,\delta} =
-\frac{n^\delta \bar h_\ell^{(2)}(ka) \bar \jmath_\ell'(nka)
- \bar h_\ell^{(2)\prime}(ka) \bar \jmath_\ell(nka)}
{n^\delta \bar h_\ell^{(1)}(ka) \bar \jmath_\ell'(nka) 
- \bar h_\ell^{(1)\prime}(ka) \bar \jmath_\ell(nka)}\,,
\end{equation}
where we have defined the Riccati-Hankel functions
$\bar \jmath_\ell(z) = z j_\ell(z)$,
$\bar h_\ell^{(1)}(z) = z h_\ell^{(1)}(z)$, and
$\bar h_\ell^{(2)}(z) = z h_\ell^{(2)}(z)$,
$\delta = \pm 1$ for the two transverse polarization channels, and
the permittivity is
\begin{equation}
\epsilon(\bm{r}) = \left\{ \begin{array}{l@{\quad}l}
n^2 & r<a \cr
1   & r>a
\end{array} \right. \,.
\end{equation}
By using smooth functions that closely approximate the step functions
in each case, we can verify that we obtain these results using our
variable phase calculation.
\end{itemize}

\subsection{Sample Calculations}

\begin{figure}[htbp]
\hfill \includegraphics[width=0.4\linewidth]{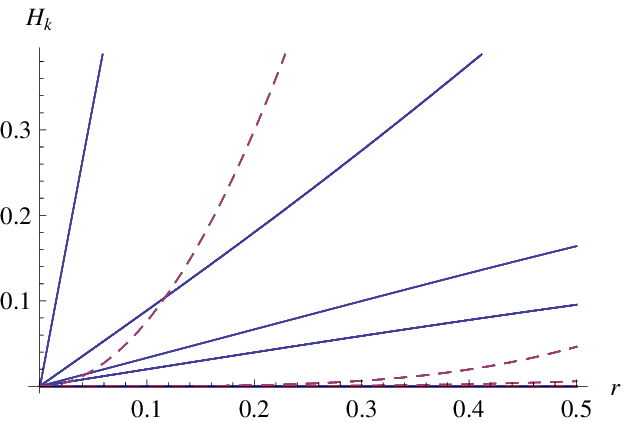} \hfill
\includegraphics[width=0.4\linewidth]{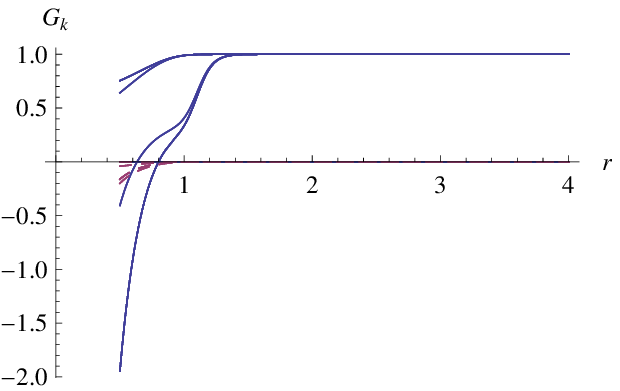} \hfill
\caption{Eigenvalues of the matrices $\hat G_k(r)$ and $\hat H_k(r)$
for $k=1$, truncated at $j_{\hbox{\tiny max}} = 2$, using the
dielectric function in Eq.~(\ref{eqn:sqpot}) with $h=4$, $w=1$, and
$s=8$.  For each eigenvalue, solid lines show the real part and dashed
lines show the imaginary part.  Taking $r_0=\frac{1}{2}$, we only
calculate $\hat H_k(r)$ for $r<r_0$ and $\hat G_k(r)$ for $r>r_0$.
}
\label{fig:gh}
\end{figure}

To illustrate the numerical advantages of the variable phase method,
we first consider a spherically symmetric example in electromagnetism,
with
\begin{equation}
\epsilon_{\ell m}(r) = \sqrt{4\pi}  \delta_{\ell 0} \delta_{m 0}
\left(1+h \, \frac{1-\tanh \left[s (r-w) \right]}{2}\right)\,.
\label{eqn:sqpot}
\end{equation}
This profile gives a smooth approximation to a dielectric ball
parameterized by height $h$, radius $w$, and edge steepness $s$.  Because
the profile is symmetric, the $S$-matrix is diagonal and degenerate in
the azimuthal quantum number $m$.  Choosing our numerical matching
point at $r_0 = \frac{w}{2}$, we integrate outward starting from 
a small radius
$r_{\hbox{\tiny small}} \ll \min\left(\frac{1}{k},w\right)$ to
obtain $\hat H_k(r)$ for $r_{\hbox{\tiny small}}<r<r_0$, and integrate
inward starting from a large radius $r_{\hbox{\tiny big}} \gg
\max\left(\frac{1}{k}, w\right)$ to obtain $\hat G_k(r)$ for
$r_{\hbox{\tiny big}}>r>r_0$.  Sample results are shown in
Fig.~\ref{fig:gh}.  We see that these functions vary smoothly in
response to the dielectric source, with trivial behavior outside the
dielectric and no oscillations.  In particular, $\hat G_k(r)$ only
becomes nontrivial when we reach values of $r$ for which the source is
no longer negligible; by choosing a moderate value of the steepness
parameter $s$, we have softened the edge of the dielectric ball in
order to highlight this transition.

\begin{figure}[htbp]
\includegraphics[width=0.4\linewidth]{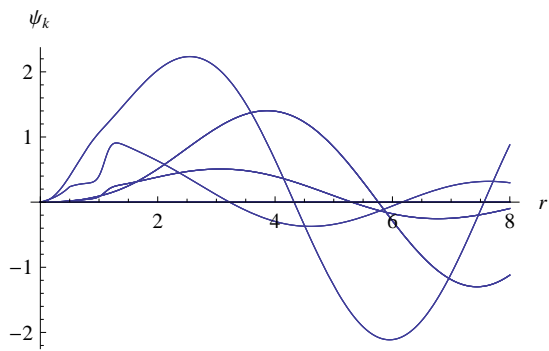}
\caption{Eigenvalues of the matrix
$\hat\psi_k^{\hbox{\tiny norm}}(r)$
for $k=1$, truncated at $j_{\hbox{\tiny max}} = 2$, using the
dielectric function in Eq.~(\ref{eqn:sqpot}) with $h=4$, $w=1$, and
$s=8$.  The two expressions in Eq.~(\ref{eqn:psinorm}) join smoothly
at $r_0 = \frac{1}{2}$.
}
\label{fig:psi}
\end{figure}

For comparison, we can reconstruct the normalized physical wavefunction
$\hat\psi_k^{\hbox{\tiny norm}}(r)$ from these results by writing
\begin{equation}
\hat\psi_k^{\hbox{\tiny norm}}(r) = 
\frac{1}{\sqrt{2\pi}} \cdot \left\{
\begin{array}{cc}
\hat G_{-k}(r) \hat W(-kr) \hat P 
\left(\widetilde{\cal W}_{-k}\right\vert_{r=r_0} \hat P
-\hat G_k(r) \hat W(kr) \hat P 
\left(\widetilde{\cal W}_k\right\vert_{r=r_0} \hat P
& \hbox{~~for $r>r_0$}
\cr
\hat W(kr)^{-1} \hat H_k(r) \hat C_k
& \hbox{~~for $r<r_0$}
\end{array}
\right.,
\label{eqn:psinorm}
\end{equation}
where $\hat P$ is the projection matrix onto the transverse modes,
the modified Wronskian $\widetilde{\cal W}_k$ is evaluated at $r=r_0$
using Eqs.~(\ref{eqn:Wtilde}) and (\ref{eqn:Wronk}), and $\hat C_k$ is
a constant matrix that matches the normalization of the two solutions,
which is obtained by setting the two expressions in
Eq.~(\ref{eqn:psinorm}) equal at $r=r_0$.  This result, shown in
Fig.~\ref{fig:psi}, displays the typical oscillations associated with
wave number $k$.  By ``factoring out'' the free contribution $\hat
W(kr)$, our method allows us to avoid these oscillations in numerical
calculations.

\begin{figure}[htbp]
\hfill \includegraphics[width=0.4\linewidth]{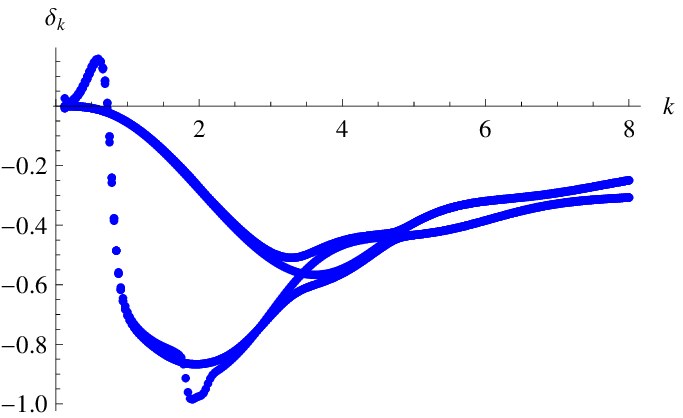}
\hfill \includegraphics[width=0.4\linewidth]{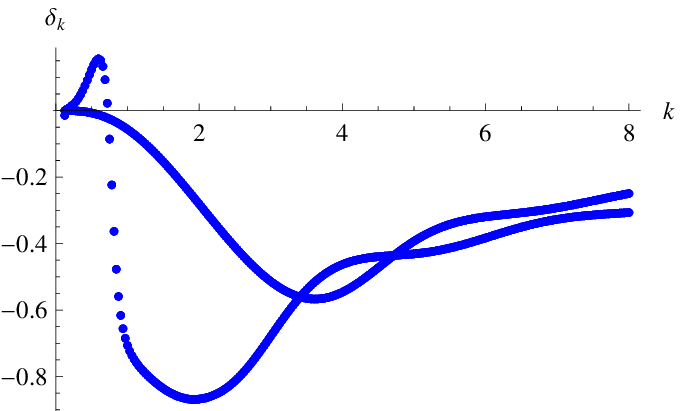} \hfill
\caption{Eigenphase shifts, given by one-half the argument of the
eigenvalues of the $S$-matrix, truncated at $j_{\hbox{\tiny max}} =
1$.  The left panel shows the case of the dielectric function given by
Eqs.~(\ref{eqn:drude}) and (\ref{eqn:drudeprofile}), with
$\lambda_p=\pi$, $\sigma_p=1$, $w=1$, and $s=8$, while the right panel
shows the result for the same $\epsilon_{00}(r)$, but with
$\epsilon_{10}(r)=0$.
}
\label{fig:S}
\end{figure}

To illustrate the $S$-matrix as a function of $k$, we consider a
dielectric with a Drude model dependence on wave number, 
\begin{equation}
\epsilon_{\ell m}(r) = \sqrt{4\pi} \, \delta_{\ell 0} \delta_{m 0}
+ \frac{(2\pi)^2}{\frac{\pi}{\sigma_p} \sqrt{-k^2}-(\lambda_p k)^2}
p_{\ell m}(r) \,,
\label{eqn:drude}
\end{equation}
where $p_{\ell m}(r)$ specifies the radial profile function for each
spherical component of the dielectric profile.  Here $\lambda_p$ is
the plasma wavelength, $\sigma_p$ is the conductivity,
and the frequency is $\omega = c\sqrt{k^2}$.  We consider a
deformed sphere using a profile given by
\begin{equation}
p_{00}(r) = \sqrt{4\pi}\frac{1-\tanh \left[s (r-w) \right]}{2} 
\hbox{\qquad and \qquad}
p_{10}(r) = \frac{1-\tanh \left[s (r-w) \right]}{2}
\,,
\label{eqn:drudeprofile}
\end{equation}
with all other $p_{\ell m}(r)$ equal to zero.  The $j=1$ eigenphase
shifts for this case, given by one-half of the argument of the
eigenvalues of the $S$-matrix, are shown in Fig.~\ref{fig:S} as
functions of $k$.  By comparing to the case where $\epsilon_{00}(r)$
is kept the same but $\epsilon_{10}(r)$ is set to zero, we see that
a nontrivial $\epsilon_{10}(r)$ mixes the polarization channels and
splits the degeneracy between $|m|=1$ and $m=0$.  As expected, these
effects vanish at small $k$, where modes have wavelengths much larger
than the length scale associated with the asymmetry, and also at large
$k$, where modes have wavelengths much smaller than the plasma
wavelength.

\section{Discussion and Future Developments}

We have developed a variable phase method to calculate the
scattering $S$-matrix for potentials in quantum mechanics and
dielectrics in electromagnetism that are localized but do not have any
particular symmetries.  The result takes the form of a matrix initial
value ODE given in terms of a spherical harmonic decomposition of the
scattering source.  By using the Wronskian, we can combine inward and
outward integration in $r$ to obtain a well-behaved numerical
computation, which remains tractable even for imaginary wave number $k$.
Finally, we have extended this approach to the electromagnetic case by
considering a modification of the Maxwell wave equation that avoids
problems associated with disentangling the transverse and longitudinal
waves.

Our high-level Mathematica code provides a transparent and
flexible high-level implementation of the methods described here, but
it is only suitable for small- to moderate-scale calculations.
Larger-scale calculations involving large numbers of partial waves
will require the use of optimized low-level parallel linear algebra
routines.  Since the ultimate problem to be solved is quite generic,
such calculations can take advantage of standard numerical packages
for matrix ODEs.

\section{Acknowledgements}

N.\ G. thanks G.\ Bimonte, J.\ Dunham,
T.\ Emig, R.\ L.\ Jaffe, M.\ Kardar, M.\
Kr\"uger, M.\ Maghrebi, M.\ Quandt, H.\ Reid, and H.\ Weigel
for helpful conversations, suggestions, and references.  A.\ F.\ and
N.\ G. were supported in part by the National Science Foundation (NSF)
through grants PHY-0855426 and PHY-1213456.

\appendix
\section{Differential Operators}
\label{appa}

Here we collect the differential operator relations needed to express
Eq.~(\ref{eqn:generalizedHelmholtz}) in the vector spherical harmonic
basis, taken from Ref.~\cite{Varshalovich:1988ye}.  In these equations
$f(r)$ is an arbitrary radial function, $Y_{j}^{m}(\theta,\phi)$ is an
ordinary spherical harmonic, and $\bm{Y}^{\ell}_{jm}(\theta,\phi)$ is
a vector spherical harmonic.
\begin{eqnarray}
\nabla \left(f(r) Y_{j}^{m}(\theta,\phi)\right)
&=& \sqrt{\frac{j}{2j+1}} \left(
\frac{d}{dr} + \frac{j+1}{r} \right)
f(r) \bm{Y}^{\ell=j-1}_{jm}(\theta,\phi)
- \sqrt{\frac{j+1}{2j+1}} \left(
\frac{d}{dr} - \frac{j}{r} \right)
f(r) \bm{Y}^{\ell=j+1}_{jm}(\theta,\phi) \cr
\nabla \cdot \left(f(r) \bm{Y}^{\ell=j+1}_{jm}(\theta,\phi)\right)
&=& -\sqrt{\frac{j+1}{2j+1}} \left(
\frac{d}{dr} + \frac{j+2}{r} \right) f(r) Y_{j}^{m}(\theta,\phi) \cr
\nabla \cdot \left(f(r) \bm{Y}^{\ell=j}_{jm}(\theta,\phi)\right) &=& 0 \cr
\nabla \cdot \left(f(r) \bm{Y}^{\ell=j-1}_{jm}(\theta,\phi)\right)
&=& \sqrt{\frac{j}{2j+1}} \left(
\frac{d}{dr} - \frac{j-1}{r} \right) f(r) Y_{j}^{m}(\theta,\phi) \cr
\nabla \times \left(f(r) \bm{Y}^{\ell=j+1}_{jm}(\theta,\phi)\right)
&=& i\sqrt{\frac{j}{2j+1}} \left(
\frac{d}{dr} + \frac{j+2}{r} \right) 
f(r) \bm{Y}^{\ell=j}_{jm}(\theta,\phi) \cr
\nabla \times \left(f(r) \bm{Y}^{\ell=j}_{jm}(\theta,\phi)\right)
&=& i\sqrt{\frac{j}{2j+1}} \left(\frac{d}{dr} - \frac{j}{r} \right)
f(r) \bm{Y}^{\ell=j+1}_{jm}(\theta,\phi) +
i\sqrt{\frac{j+1}{2j+1}} \left(\frac{d}{dr} + \frac{j+1}{r} \right)
f(r) \bm{Y}^{\ell=j-1}_{jm}(\theta,\phi) \cr
\nabla \times \left(f(r) \bm{Y}^{\ell=j-1}_{jm}(\theta,\phi)\right)
&=& i\sqrt{\frac{j+1}{2j+1}} \left(
\frac{d}{dr} - \frac{j-1}{r} \right) f(r) \bm{Y}^{\ell=j}_{jm}(\theta,\phi)
\end{eqnarray}

\section{Free Green's Functions and Plane Wave Expansions}

Throughout this paper we have considered scattering in a spherical
partial wave basis.  For both Casimir calculations and traditional
scattering problems, it is helpful to be able to convert these results
to a plane wave basis.  The key tools in this conversion are the
expansion of a plane wave and the expansion of the free Green's
function in terms of free spherical waves.  Again drawing on 
Ref.\ \cite{Varshalovich:1988ye}, we collect those expansions here.
For scalar scattering we have the well-known results
\begin{equation}
e^{i\bm{k}\cdot \bm{r}} = 4 \pi\sum_{\ell m} i^\ell j_\ell(kr) 
Y_\ell^m(\theta_k, \phi_k)^\ast
Y_\ell^m(\theta, \phi)\,,
\end{equation}
where $\theta_k$ and $\phi_k$ are the angles of $\hat{\bm{k}}$ in
spherical coordinates, and
\begin{equation}
{\cal G}_0(\bm{r},\bm{r'},k) = ik \sum_{\ell m}
j_\ell(kr_<) h^{(1)}_\ell(kr_>)
Y_\ell^m(\theta', \phi')^\ast Y_\ell^m(\theta, \phi)
\,,
\end{equation}
where $r_<$ ($r_>$) is the smaller (larger) of $r=|\bm{r}|$ and
$r'=|\bm{r}'|$.  For vector waves, the expansion of a plane wave with
polarization $\bm{\xi}$ becomes
\begin{equation}
\bm{\xi} e^{i \bm{k} \cdot \bm{r}} =
4\pi \sum_{\ell j m} i^\ell
\left(\bm{\xi} \cdot \bm{Y}^{\ell}_{jm}(\theta_k, \phi_k)^\ast\right)
j_\ell(kr) \bm{Y}^{\ell}_{jm}(\theta, \phi)\,,
\end{equation}
while the expansion of the free dyadic Green's function is
\begin{equation}
\mathbb{G}(\bm{r}_1, \bm{r}_2, k) = 
ik \sum_{\ell j m} j_\ell(k r_<) h^{(1)}_\ell(k r_>)
\bm{Y}^{\ell}_{jm}(\theta_1, \phi_1)^\ast \otimes
\bm{Y}^{\ell}_{jm}(\theta_2, \phi_2)\,.
\end{equation}
We can also express these results in terms of transverse and
longitudinal vector spherical harmonics.  For the decomposition of a
vector plane wave, we define
\begin{eqnarray}
\bm{Y}^M_{j m}(\theta, \phi) &=& 
\bm{Y}^{\ell=j}_{j m}(\theta, \phi) \cr
\bm{Y}^N_{j m}(\theta, \phi) &=& 
\sqrt{\frac{j+1}{2j+1}} \bm{Y}^{\ell=j-1}_{j m}(\theta, \phi) +
\sqrt{\frac{j}{2j+1}} \bm{Y}^{\ell=j+1}_{j m}(\theta, \phi) \,,
\end{eqnarray}
for $j=1,2,3\ldots$, and
\begin{equation}
\bm{Y}^L_{j m}(\theta, \phi) =
\sqrt{\frac{j}{2j+1}} \bm{Y}^{\ell=j-1}_{j m}(\theta, \phi) -
\sqrt{\frac{j+1}{2j+1}} \bm{Y}^{\ell=j+1}_{j m}(\theta, \phi)
\end{equation}
where $j=0,1,2,3\ldots$.  (Note that for $j=0$, the unphysical term
with $\ell=-1$ is multiplied by zero.)  
Similarly, we consider the free transverse modes in
Eqs.~(\ref{eqn:transverse}) along with the free longitudinal mode,
given by
\begin{equation}
\bm{L}_{jm,k}(r,\theta,\phi) = 
\sqrt{\frac{j}{2j+1}} z_{j-1}(kr) 
\bm{Y}^{\ell=j-1}_{jm}(\theta, \phi)
+\sqrt{\frac{j+1}{2j+1}} z_{j+1}(kr) 
\bm{Y}^{\ell=j+1}_{jm}(\theta, \phi)
\label{eqn:longitudinal}
\end{equation}
for $j=0,1,2\ldots$.

For the decomposition of a plane wave, we then have
\begin{equation}
\bm{\xi} e^{i \bm{k} \cdot \bm{r}} =
4\pi \sum_{\bm{\chi} j m} i^{j+\sigma}
\left(\bm{\xi} \cdot \bm{Y}^{\chi}_{jm}(\theta_k, \phi_k)^\ast\right)
\bm{\chi}^{\rm reg}_{jm,k}(r,\theta,\phi)\,,
\end{equation}
where $\sigma = 0,1,-1$ for $\bm{\chi} = \bm{M},\bm{N},\bm{L}$
respectively, and for the free dyadic Green's function we have
\begin{equation}
\mathbb{G}(\bm{r}_1, \bm{r}_2, k) = 
ik \sum_{\bm{\chi} j m}
\bm{\chi}^{\rm reg}_{jm,k}(\bm{r}_<)^\ast \otimes
\bm{\chi}^{\rm out}_{jm,k}(\bm{r}_>)\,.
\end{equation}
again for $\bm{\chi} = \bm{M},\bm{N},\bm{L}$.  Here the
regular solution is given by taking
$z_\ell(k r) = j_\ell(k r)$ in Eqs.~(\ref{eqn:transverse}) and
(\ref{eqn:longitudinal}), while the outgoing
solution has $z_\ell(k r) = h^{(1)}_\ell(k r)$.

\bibliographystyle{apsrev}
\bibliography{article}

\end{document}